\documentclass{article}
\usepackage{smc2026}
\usepackage[T1]{fontenc}
\usepackage{booktabs}
\usepackage{subcaption}
\usepackage{multirow}
\usepackage{musicography}

\DeclareMathOperator*{\argmax}{arg\,max}

\providecommand{\capstartfalse}{}
\providecommand{\capstarttrue}{}

\def\papertitle{Diachronic Modeling of Tonal Coherence on the Tonnetz Across Classical and Popular Repertoires\thanks{Supplementary materials, code and data available at \url{https://github.com/alunxu/tonal-coherence}}}

\author[1]{\mbox{\firstname{Weilun}\lastname{Xu}}}
\author[2]{\mbox{\firstname{Edward}\lastname{Hall}\orcid{0000-0002-8268-5052}}}
\author[2]{\mbox{\firstname{Martin}\lastname{Rohrmeier}\orcid{0000-0002-4323-7257}}}

\affil[1]{\department{School of Computer and Communication Sciences}\institution{École Polytechnique Fédérale de Lausanne}\country{Switzerland}\affiliationtype{University}}
\affil[2]{\department{Digital and Cognitive Musicology Lab}\institution{École Polytechnique Fédérale de Lausanne}\country{Switzerland}\affiliationtype{University}}

\completesetup

\title{\papertitle}

\begin{document}
\capstartfalse
\maketitle
\capstarttrue

\begin{abstract}
How do different musical traditions achieve tonal coherence? Most computational measures to date have analysed tonal coherence in terms of a single dimension, whereas a multi-dimensional analyses have not been sufficiently explored. We propose a new model drawing on the concept of the Tonnetz---we define two partially independent measures: \emph{tonal focus}, the concentration of pitch content near a tonal center; and \emph{tonal connection}, the degree to which pitch content reflects structured intervallic pathways back to that center. Analyzing over 2,800 pieces from Western classical and popular traditions, we find that these traditions occupy overlapping yet distinguishable regions of the two-dimensional space. Popular music shows higher tonal focus, while classical music exhibits higher tonal connection. Our complementary measures ground the differences between different tonal styles in quantitative evidence, and offer interpretable dimensions for computational music analysis and controllable generation.
\end{abstract}

\section{Introduction}

Tonal coherence---the perception that pitch relationships follow a meaningful organization, directed towards a ``home'' tonic pitch---is fundamental to Western music theory. This formulation of tonal coherence could also be conceived of in terms of ``centricity'' such that a pitch can function as a center even ``without the support of functional harmonic relations.''\cite{straus2016} 
Computational musicology has made significant progress in investigating tonality and tonal structure \cite{meredith2016}; for instance, models have captured how pitch distributions relate to key centers \cite{temperley2007music,krumhansl1990,huron2006,quinn2017}, and how pitch distributions have changed over time \cite{serra2012measuring,yust2019stylistic, harasim2021exploring,moss2024computational}. 
Up to the present, many approaches to tonality are effectively grounded in the unidimensional representation of pitch class distributions \cite{huron2006}.
On the other hand, aspects of tonality and tonal coherence have not yet been explored in comparable depth with multi-dimensional representations of pitch-class distributions, such as the Tonnetz.


\begin{figure}[!t]
\centering
\includegraphics[width=\columnwidth]{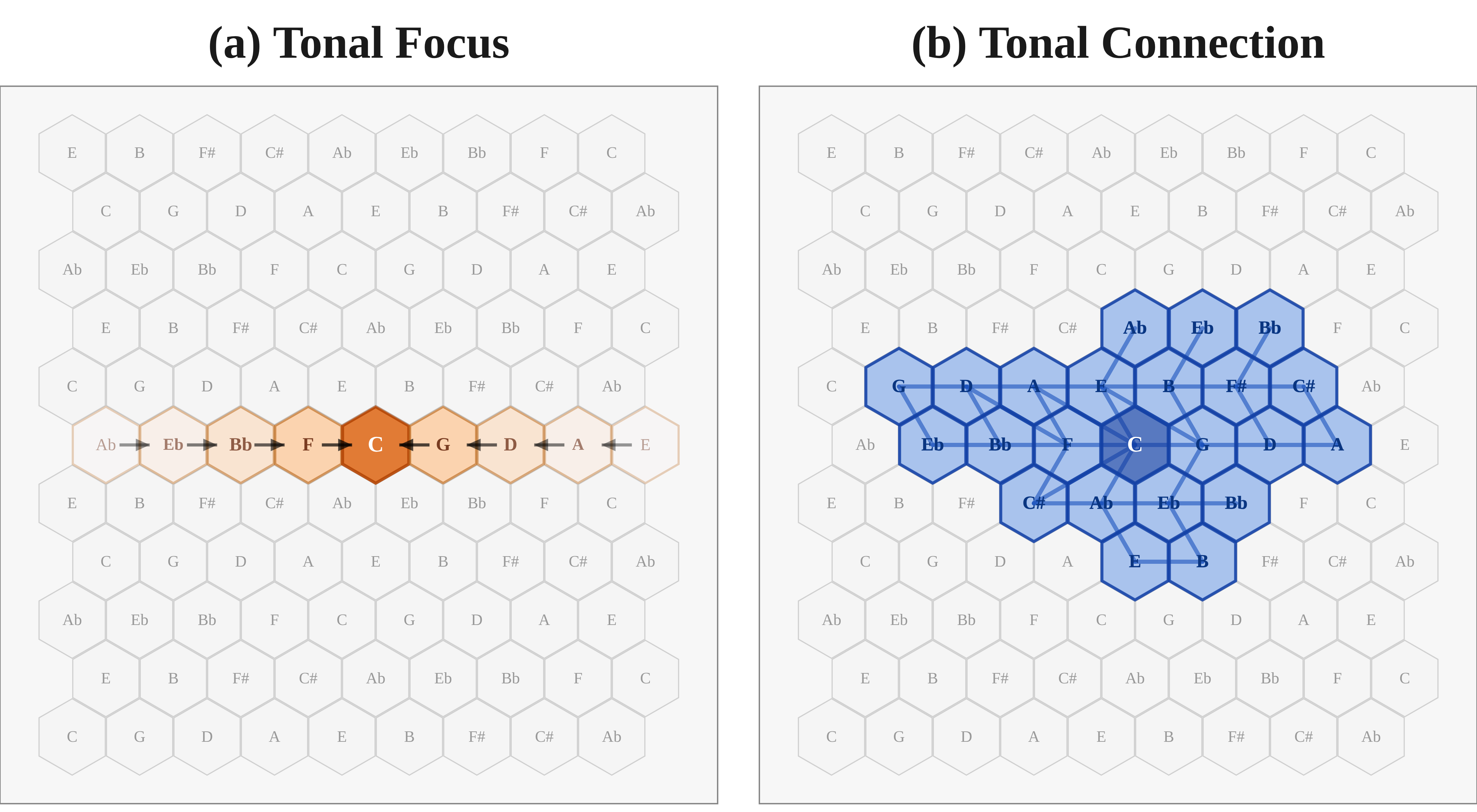}
\caption{\textbf{Two dimensions of tonal organization.} (a) \emph{Tonal focus} (gravitational centering) along the line-of-fifths: Pitch content is concentrated within a small diatonic region around the tonic, showing strong gravitational pull but limited exploration. (b) \emph{Tonal connection} (structured exploration) across the Tonnetz lattice: Pitch content systematically explores distant tonal regions through structured intervallic pathways, maintaining coherence despite potentially lower concentration near the tonic.}
\label{fig:conceptual}
\end{figure}

We propose that tonality and, specifically, the aspect of tonal coherence can be well modeled using two measures drawing on the music theoretical concept of the \textit{Tonnetz} (illustrated in Figure~\ref{fig:conceptual}). \emph{\textbf{Tonal focus}} captures how tightly pitch content concentrates near the tonic. \emph{\textbf{Tonal connection}} captures the degree to which a piece's pitch content reflects systematic intervallic relationships---construing individual pitches to be arrived at by traversing the Tonnetz 
%
rather than emerging independently of any context.
These measures are partially independent. The pitch content of a piece can stay close to the tonal center without systematically connecting pitches through intervallic structures (low connection, high focus), or conversely, provide extensive harmonic exploration while maintaining clear pathways to the center (high connection, moderate focus). We characterize focus through concentration within the diatonic region of the line-of-fifths, and connection through the Tonal Diffusion Model's path-length parameter and interval weight structure \cite{lieck2020tonal}.

\section{Background}

\subsection{Theoretical Approaches to Tonal Coherence}

Music theory addresses, in different ways, the question of how tonal coherence arises. Possibly the first attempt to codify tonal functional harmony was by Rameau \cite{rameau1722}. In the early 20th century, Schenker developed a theory and analytical framework of tonal music based on the concept of hierarchical prolongation \cite{schenker1935,cadwallader1998}. Around the same time, composer and theorist Schoenberg proposed a concept of extended tonality resulting from stretching the boundaries of traditional tonal harmony to accommodate enhanced musical expression \cite{schoenberg1969}. The Tonnetz was originally conceived to model musical temperament \cite{euler1739} and was later adapted as an algebraic tonal space of pitch-class and chord relations \cite{oettingen1866, hostinsky1879, riemann1893, cohn1997, lerdahl2001}. More recent Neo-Riemannian Theory explores the algebraic dual space of the hexagonal Tonnetz as a model of triadic chord transformations \cite{cohn1997}. These transformational accounts of harmony \cite{lewin1987generalized} are capable of expressing harmonic relations as found in the "extended common practice" (as coined by Tymoczko \cite{tymoczko2011}), evolving throughout the 19th and 20th centuries. The recent approaches of Tonfeld theory \cite{haas2004,rohrmeier2021} similarly propose to model extended tonality in terms of structures derivable from the Tonnetz.

Scholars of popular music have argued tonal coherence can be achieved differently across genres. Everett \cite{everett2004rock} notes that voice-leading in rock is often ``severely compromised''. Biamonte \cite{biamonte2010triadic} shows rock progressions achieve coherence through modal centricity rather than leading-tone resolution. Tagg \cite{tagg2014everyday} argues that loop-based music achieves coherence through ``assertional'' centering. These observations suggest different musical genres may weight the two measures differently, but direct quantitative comparison has been lacking.

\subsection{Computational Models of Tonal Structure}

Geometric representations of tonal space (the Tonnetz, see also \cite{chew2001modeling} spiral array, \cite{lerdahl2001tonal} pitch space) established that music-theoretic intuitions could be captured in spatial models. The \textit{Tonal Diffusion Model} (TDM) \cite{lieck2020tonal} advanced this by providing a generative account. Pitch-class distributions arise from random walks using fifths and thirds, with parameters capturing traversal extent ($\lambda$) and interval preferences (weight vector $\mathbf{w}$). The TDM has characterized historical trends in Western classical music \cite{moss2024computational}. We extend it by decomposing its parameters into interpretable dimensions and applying it to cross-genre comparison. We adopt the TDM over alternatives \cite{chew2001modeling,bernardes2016tiv,lerdahl2001tonal} because it uniquely provides both a traversal parameter and interval weight structure that directly operationalize our two-dimensional framework.

Information-theoretic measures provide complementary characterizations. Yust's DFT analysis \cite{yust2019stylistic} revealed trends in diatonicity. Wei{\ss} et al. \cite{weiss2019investigating} traced stylistic evolution across 300 years. Harasim et al. \cite{harasim2021exploring} used Bayesian modeling to show how modal structure shifted across historical epochs. Network-based approaches \cite{nardelli2022towards,bono2025network} have characterized complexity through entropy and connectivity. These methods typically treat complexity as unidimensional; our work shows that apparent complexity differences may reflect qualitatively different strategies.

\subsection{Cross-Tradition Comparison}

Annotated corpora have enabled empirical comparison across traditions. De Clercq and Temperley's rock corpus \cite{declercq2011}  revealed the prevalence of I, IV, and V chords. Temperley \cite{temperley2018} further formalized rock harmony through probabilistic modeling. Serr\`a et al. \cite{serra2012measuring} documented pitch restriction in contemporary popular music. Di Marco et al. \cite{dimarco2025decoding} compared complexity across genres using network measures. These studies interpret cross-genre differences as degree variations along a single axis (more/less complex, more/less chromatic) rather than as qualitatively different strategies---a limitation our two-dimensional framework directly addresses.

Using the Distant Listening Corpus \cite{hentschel2025dlc} for classical music (1680--1920) and the Lakh MIDI Dataset \cite{raffel2016lmd} for popular music (1950--2020), we derive four hypotheses: (1) tonal focus and connection are only moderately correlated; (2) classical music exhibits higher fifth dominance with more differentiated interval weights; (3) classical music shows higher mean $\lambda$ with lower variance; and (4) the classical tradition shows a historical trajectory from Baroque constraint toward early 20th-century diversity.

\section{Methods}

\subsection{Datasets}

We analyze two complementary corpora representing distinct traditions within Western tonal music.

\textbf{Classical corpus.}\quad We use the Distant Listening Corpus (DLC)\footnote{\url{https://github.com/DCMLab/distant_listening_corpus}} \cite{hentschel2025dlc}, comprising 1,326 expert-annotated and cross-reviewed pieces by 36 composers from the common-practice period (1680--1920). The DLC was developed at EPFL's Digital and Cognitive Musicology Lab (DCML), and provides symbolically encoded scores following the DCML standard. 

\textbf{Popular corpus.}\quad We draw a genre-balanced sample from the Lakh MIDI Dataset (LMD) \cite{raffel2016lmd} across 12 genres.\footnote{Rock, Pop, Country, Electronic, R\&B, Metal, Rap, Latin, Reggae, Folk, World, Punk.} We excluded pieces with fewer than 5 unique pitch classes, pitch entropy outside the 1.5--3.2\,bit range (the upper bound removes unrealistically flat distributions from noisy MIDI transcriptions), any single pitch class exceeding 50\% of duration, or tonal focus below 0.3 (indicating failure to concentrate on any tonal center); genre filters removed classical, jazz, blues, and new age (to avoid overlap with the classical corpus and genres whose tonal practices fall outside the popular-music scope of this study). After all filtering, 1,569 pieces remain. Since MIDI key signatures are unreliable, we estimate tonal centers algorithmically.

The corpora differ in annotation quality---DLC provides expert harmonic analysis while LMD relies on user-generated transcriptions---which we address through robustness analyses.\footnote{See supplementary material, Appendix~B, \url{https://github.com/alunxu/tonal-coherence}}

\subsection{Tonal Representation}

We represent pitch content using 35-dimensional vectors ordered along the line-of-fifths, from F\musDoubleFlat\ (index 0) through A\musDoubleSharp\ (index 34), with C at the center (index 17). This representation preserves enharmonic distinctions and directly encodes the fifth relationships central to our framework. For each piece, we construct a normalized distribution $\mathbf{d} \in \mathbb{R}^{35}$ where each $d_i$ is the proportion of total sounding duration at position $i$. This static, piece-level representation captures \emph{what} pitches occur but not \emph{when}.

The classical corpus provides explicit pitch spellings (tonal pitch class values) in its symbolic score annotations. For the popular corpus, which consists of MIDI files without enharmonic information, we infer spellings using the \texttt{partitura} library \cite{cancino2018partitura}, which applies key estimation and voice-leading heuristics. While this introduces potential spelling noise, a 12-dimensional reanalysis confirmed that main findings hold under chromatic collapsing.\footnote{See supplementary material, Appendix~B.}

\subsection{Tonal Center Estimation}
\label{sec:tonal_center}

Our measures require specification of a tonal center $c$ for each piece.

\textbf{Classical corpus.}\quad We use expert-annotated global keys from the DLC metadata, which provide a single key label per movement as determined by analysts.

\textbf{Popular corpus.}\quad We estimate tonal centers using the well-established Krumhansl-Schmuckler algorithm \cite{krumhansl1990}, which computes Pearson correlations between each piece's chromatic pitch-class profile and theoretical major and minor key profiles, selecting the best-matching key. The estimated key is then expressed as a line-of-fifths index for use with the TDM.

Both corpora assume a single global tonal center---appropriate for most popular music but problematic for modulating classical works. A sonata spending substantial time in the dominant will show pitch content ``off-center'' relative to its global tonic. This limitation likely inflates observed genre differences; classical music's measured focus may partly reflect modulation rather than within-key chromaticism. Modulation-stratified analysis confirmed that weight structure metrics remain stable across modulation rates.

\begin{figure*}[!t]
\centering
\includegraphics[width=\textwidth]{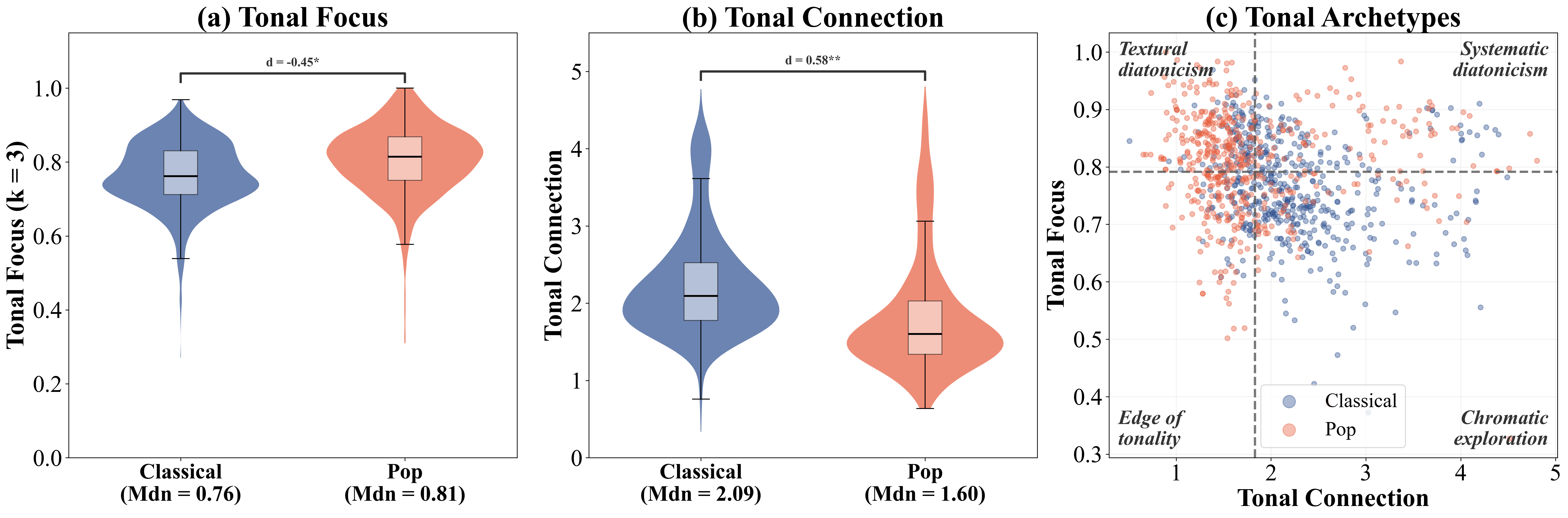}
\caption{\textbf{Two dimensions of tonal organization.} (a) \emph{Tonal focus} ($k$: window half-width on the line-of-fifths): popular music concentrates more pitch content near the tonic. (b) \emph{Tonal connection} ($\lambda$: TDM path-length parameter): classical music shows higher values with notably lower variance. (c) \emph{Tonal archetypes}: the two-dimensional space reveals four characteristic regions defined by the combined median of each dimension. All effect sizes $d$ are Cohen's $d$ (pooled SD); $^{*}|d|{\geq}0.2$, $^{**}{\geq}0.5$, $^{***}{\geq}0.8$.}
\label{fig:main}
\end{figure*}

\subsection{Operationalizing the Two Dimensions}

Our theoretical framework proposes two partially independent mechanisms that contribute to tonal coherence. We operationalize both through complementary measures, \textit{tonal focus} and \textit{tonal connection}.

\subsubsection{Tonal Focus}

Tonal focus captures how tightly pitch content $d$ clusters around the tonic center, $c$. We measure the proportion of pitch content within $\pm k$ positions on the line-of-fifths:
\begin{equation}
\text{\textbf{Tonal Focus}}_k = \sum_{i=c-k}^{c+k} d_i
\end{equation}

We set $k=3$ as our primary threshold. This window captures concentration within a diatonic-scale-sized region around the tonic, although it does not perfectly align with the key's own diatonic collection; a major key's diatonic pitches span positions $-1$ (subdominant) to $+5$ (leading tone) on the line-of-fifths, an asymmetric range reflecting the leading tone's greater distance from the tonic. Our symmetric window instead treats sharp-side and flat-side deviations equally, measuring how tightly pitch content clusters near the tonal center regardless of mode. Values approaching 1.0 indicate concentration within this central region; lower values indicate chromatic extension in either direction. 

This global-key measure may underestimate focus in pieces with extensive modulation, as pitch content in secondary keys will appear displaced from the primary tonic on the line-of-fifths. Since classical music modulates more frequently than popular music, this could systematically depress classical focus values. We assess this potential confound through modulation stratification.\footnote{See supplementary material, Appendix~B.}

\subsubsection{Tonal Connection}

Tonal connection captures the extent to which a piece's pitch distribution is consistent with systematic intervallic relationships to the tonic. We quantify this using the Tonal Diffusion Model (TDM) \cite{lieck2020tonal}, which models pitch distributions as arising from random walks on the Tonnetz.

The TDM assumes pitches are generated by walks using six primary intervals: perfect fifths ($\pm 1$ step on the line-of-fifths); major thirds ($\pm 4$ steps); and minor thirds ($\mp 3$ steps, a minor third corresponds to moving 3 positions in the flat direction on the line-of-fifths). The number of steps $n$ follows a Poisson distribution with mean $\lambda$, and each step uses interval $j$ with probability $w_j$. Given an observed pitch distribution $\mathbf{d}$ (where $d_i$ is the proportion of duration at position $i$) and tonal center $c$, the model parameters are estimated via maximum likelihood,
\begin{equation}
\{\lambda^*, \mathbf{w}^*\} = \argmax_{\lambda, \mathbf{w}} \sum_{i=0}^{34} d_i \log P_{\text{model}}(i \mid \lambda, \mathbf{w}, c),
\end{equation}
where $P_{\text{model}}(i \mid \lambda, \mathbf{w}, c)$ is the probability of reaching position $i$ via random walks starting from $c$.

We define tonal connection as the fitted path-length parameter:
\begin{equation}
\text{\textbf{Tonal Connection}} = \lambda^*.
\end{equation}
Higher values indicate more extensive traversal of tonal space---reaching chromatically distant pitch classes while maintaining systematic connection through intermediate steps. Note that this measure captures distributional evidence of intervallic organization, not temporal voice-leading directly.

However, $\lambda$ alone captures \emph{how far} a piece explores tonal space, not \emph{how} it gets there. Two pieces with the same $\lambda$ could differ fundamentally in strategy, one navigating primarily through fifths (the backbone of tonal harmony) and the other distributing movement equally across fifths and thirds. To distinguish these strategies, we derive three summary statistics from the fitted interval weights $\mathbf{w}^* = [w_{-4}, w_{-3}, w_{-1}, w_{+1}, w_{+3}, w_{+4}]$:

\textbf{Fifth dominance.}\quad $(w_{-1}\!+\!w_{+1}) / {\textstyle\sum_j} w_j$ measures the proportion of intervallic weight allocated to perfect fifths. A value near 1.0 indicates that tonal space is navigated almost exclusively along the circle of fifths---the principal axis of tonal harmony since Rameau; lower values indicate greater reliance on third-based movement (characteristic of chromatic or Neo-Riemannian voice-leading).

\textbf{Weight entropy.}\quad $-{\textstyle\sum_j} w_j \log w_j$ captures the diversity of interval usage. Low entropy indicates a clear preference for one or two intervals (a ``preferred route'' through tonal space); high entropy indicates that all six intervals contribute comparably, suggesting less hierarchically structured exploration.

\textbf{Weight kurtosis.}\quad The excess kurtosis (Fisher) of the six-element weight vector. Near-zero weight kurtosis indicates a graded hierarchy---one interval dominates while others play differentiated subordinate roles (e.g., fifths primary, thirds secondary). Negative (platykurtic) values indicate a flatter distribution where no single interval is strongly preferred.

Together, these metrics decompose tonal connection into its \emph{extent} ($\lambda$) and \emph{character} (fifth dominance, weight entropy, weight kurtosis). High tonal connection, in the sense relevant to our framework, is indicated by extensive traversal (high $\lambda$) combined with structured interval preferences (high fifth dominance, near-zero weight kurtosis)---a signature we expect to differentiate classical from popular music.

\section{Results}

We present our results in the following order.
First, we establish that focus and connection are partially independent dimensions separating the classical and pop traditions (using the DLC and LMD, respectively), we then characterize the tonal archetypes that emerge at the extremes, examine the interval weight structures underlying each tradition's strategy, and trace within-tradition variation across stylistic eras.

\subsection{Two Dimensions of Tonal Organization}

Figure~\ref{fig:main} shows the distribution of tonal focus and tonal connection metrics for both datasets. These metrics form partially independent dimensions, along which the two traditions separate. Classical and popular music cluster in distinct regions of the two-dimensional space (Figure~\ref{fig:main}c), this suggests that coherence arises through multiple organizational strategies rather than varying along a single axis.

The correlation between tonal connection and tonal focus is moderate and negative ($r = -0.20$ in classical, $r = -0.06$ in popular music), suggesting a soft trade-off---pieces with more extensive tonal exploration tend to have lower focus, but the vast majority of variance in each dimension is independent. Critically, all quadrants of the space are occupied to some extent by both datasets. Some classical pieces have both high focus \emph{and} high connection (in-common with the pop dataset), while some popular pieces approach classical levels of tonal connection. This confirms that position on one dimension does not determine position on the other.

Popular music concentrates more pitch content near the tonic (Figure~\ref{fig:main}a), consistent with coherence through diatonic concentration rather than systematic intervallic exploration. Classical music's lower focus reflects its broader chromatic vocabulary, including modulation and chromatic voice-leading.

Classical pieces exhibit substantially higher tonal connection with narrower spread (Figure~\ref{fig:main}b), reflecting a more uniform organizational strategy consistent with voice-leading principles \cite{cohn1997}. Within the popular corpus, the 12 genres spread across the space rather than clustering in a single region, confirming that aggregating across genres masks meaningful organizational differences.

\subsection{Tonal Archetypes}

The dashed lines in Figure~\ref{fig:main}c mark the combined median of each dimension across both corpora, defining four regions on an absolute scale. The two datasets distribute asymmetrically across these regions, revealing their preferred coherence strategies. We characterize these regions as follows:

\textbf{Chromatic exploration} (high connection / low focus) is the classical tradition's dominant archetype, containing nearly half of all classical pieces (47\%) but only 15\% of popular pieces. Wagner's \emph{Tristan} Prelude and Liszt's \emph{Après une lecture du Dante} exemplify this ``transformational'' route to coherence \cite{cohn2012audacious}, navigating distant tonal regions through structured intervallic pathways while maintaining low diatonic concentration.

\textbf{Textural diatonicism} (low connection / high focus) is the mirror image. It is the dominant archetype for popular music (41\%) but representing only 15\% of classical pieces. This region is consistent with ``assertional'' tonality \cite{tagg2014everyday}---three-chord progressions and loop-based structures establish coherence through gravitational centering and metric emphasis rather than systematic voice-leading.

\textbf{Systematic diatonicism} (high connection / high focus) combines both forms of coherence. Bach's French Suite No.\ 5 and Schumann's \emph{Kinderszenen} No.\ 7 exemplify this region, which is roughly equally represented across traditions (23\% classical, 18\% popular), suggesting that maximal tonal organization is an available strategy regardless of tradition.

\textbf{Edge of tonality} (low connection / low focus) represents music lacking both connection and focus. Popular music shows greater representation here (26\% vs.\ 14\%), consistent with genres that achieve coherence through non-pitch means (rhythm, timbre, texture, production).

This asymmetry---classical music concentrating in the high-connection half of the space, popular music in the high-focus half---directly reflects the core theoretical distinction between transformational and functional approaches to tonal organization.

\subsection{Dimension Characterization}
\label{sec:dimensions}

\begin{figure}[!t]
\centering
\includegraphics[width=\columnwidth]{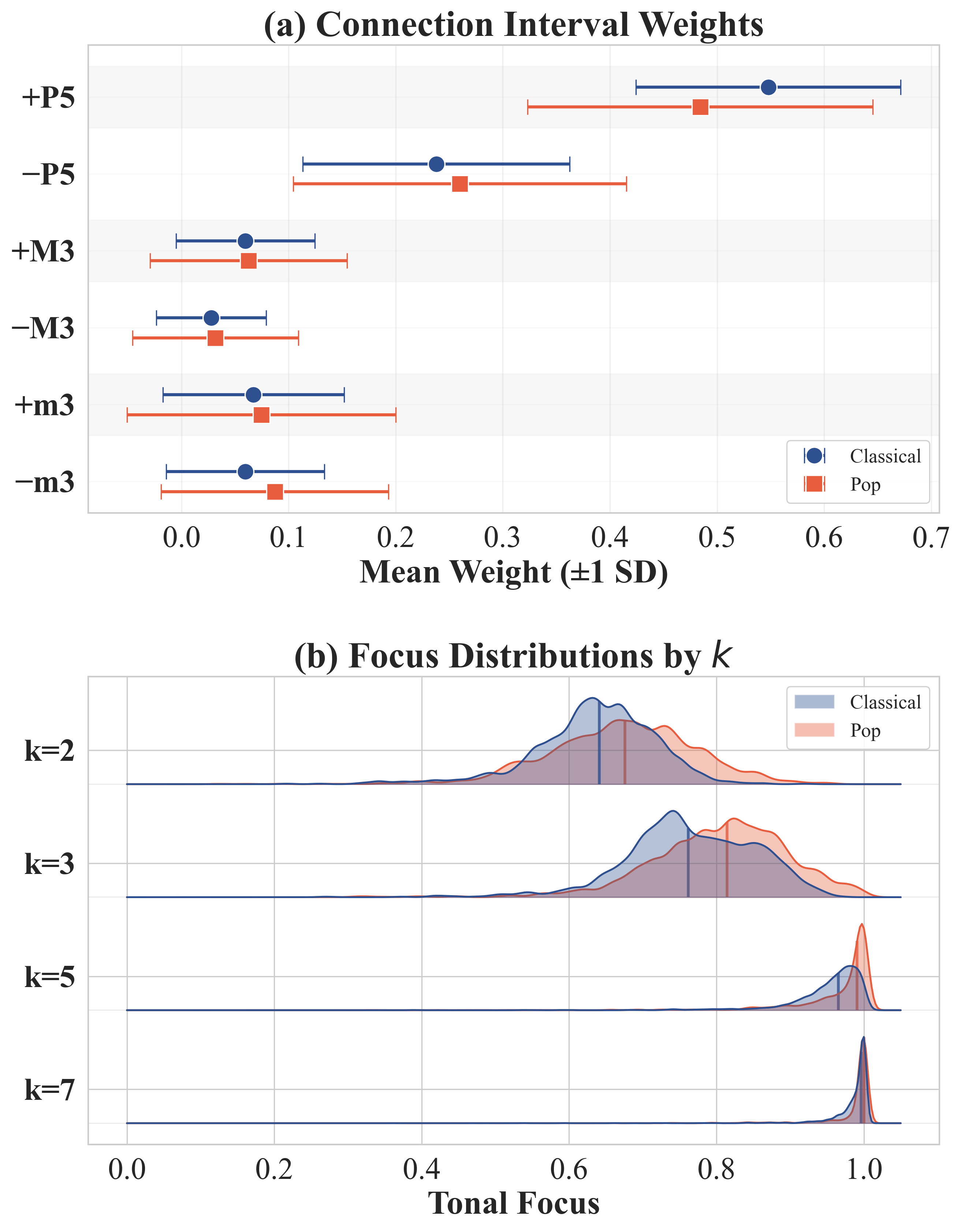}
\caption{\textbf{Characterizing the two dimensions.} (a) Mean weight profiles across the six TDM intervals with $\pm$1 SD error bars. Classical music allocates substantially more weight to perfect fifths (fifth dominance $d = 0.31$; weight kurtosis $d = 0.63$). (b) Ridge plots of tonal focus distributions from $k=2$ to $k=7$. At $k=3$ (primary threshold), classical music is unimodal while popular music is broader and bimodal; both converge toward 1.0 at large $k$.}
\label{fig:dimensions}
\end{figure}

To confirm that classical music's higher tonal connection reflects \emph{systematic} exploration rather than undirected chromatic wandering, we analyzed the interval weight vectors of the fitted TDM (Figure~\ref{fig:dimensions}a). Classical music allocates approximately 79\% of intervallic weight to perfect fifths compared to 74\% in popular music, consistent with the centrality of fifth-relations in both Schenkerian prolongation and Neo-Riemannian transformations \cite{cohn1997}, where the perfect fifth serves as the primary axis of tonal organization.

Beyond this dominance, the weight distributions differ in shape. Classical music exhibits near-zero excess kurtosis ($-0.09$), indicating a graded hierarchy where the fifth dominates while other intervals play differentiated subordinate roles, whereas popular music shows strongly negative kurtosis ($-0.56$, platykurtic), indicating flatter distributions without clear intervallic preferences. Weight entropy shows only a negligible difference, suggesting that the traditions use similar overall interval diversity but structure it differently---hierarchically in classical music versus uniformly in popular music.

This pattern---structured fifth dominance with differentiated subordinate weights---characterizes the ``transformational'' approach to coherence, using systematic navigation of tonal space through theory-canonical intervals. Popular music's flatter profile suggests coherence is achieved through other methods, such as metric emphasis and repetition, rather than hierarchical intervallic structure.

\begin{figure*}[!t]
\centering
\includegraphics[width=1.2\columnwidth]{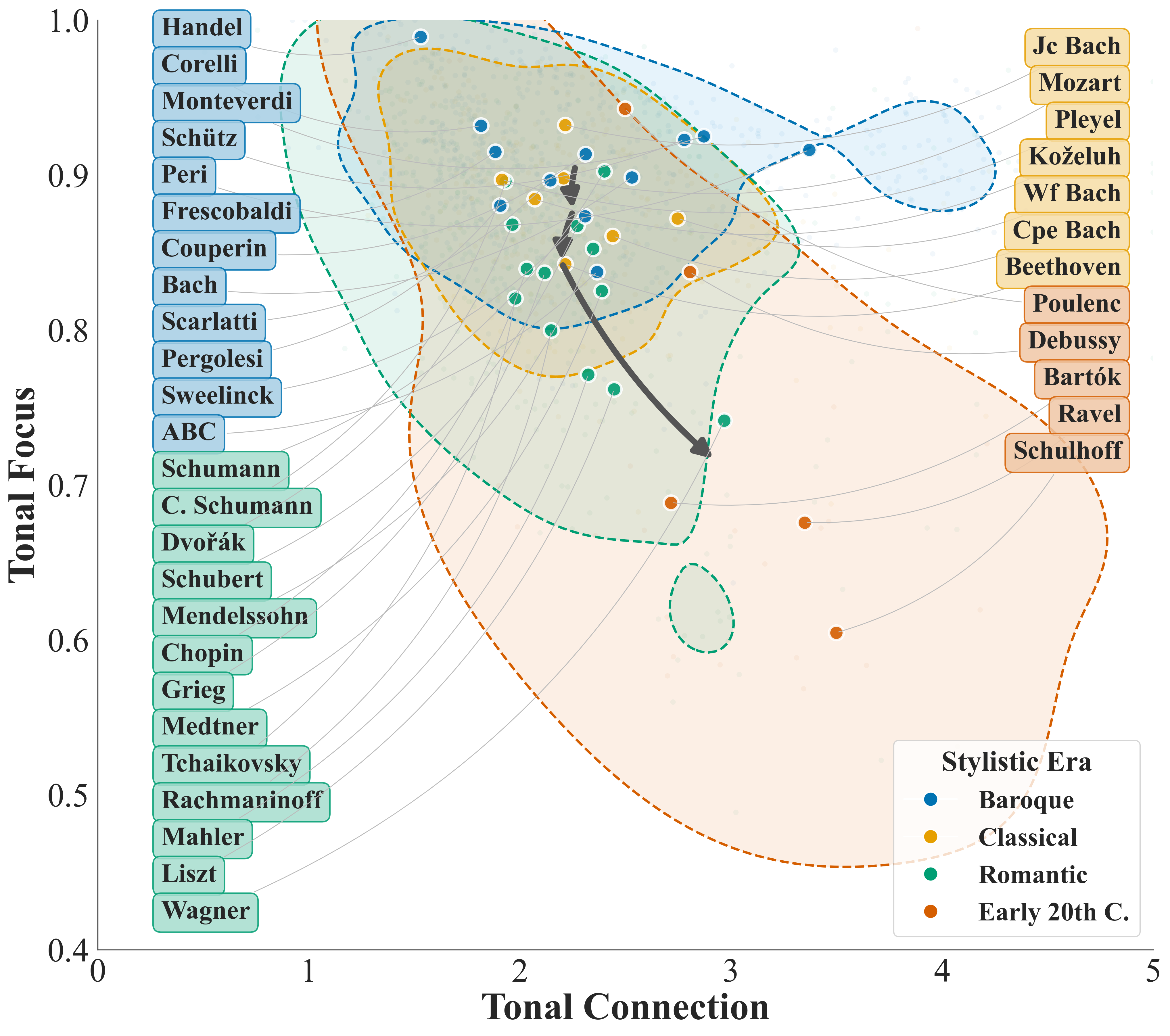}
\caption{\textbf{Historical trajectory in the two-dimensional space.} Each point represents a composer's average position, colored by stylistic era; dashed contours show KDE density regions for each era. The arrow traces the centroid path from Baroque through the early 20th century, revealing a progressive expansion from tight, high-focus clustering toward greater dispersion in both dimensions. The small disconnected Romantic contour reflects the density tail of low-focus outliers such as Wagner and Liszt.}
\label{fig:trajectory}
\end{figure*}

Within the classical tradition, composer-level analysis reveals a historical trajectory from a more constrained Baroque practice toward the greater stylistic pluralism of the late Romantic and early 20\textsuperscript{th} century eras (Figure~\ref{fig:trajectory}). Baroque composers cluster tightly in the upper region (high tonal focus with moderate tonal connection), consistent with a strong diatonic orientation and relatively constrained harmonic vocabulary. The Classical era maintains high focus but displays a shift toward higher connection, consistent with the expanding tonal palette (e.g., as used in sonata-form developments). Romantic composers spread across a much wider region, with Wagner and Mahler pulling toward lower focus while Chopin and Schumann remain closer to Classical norms. By early 20th century, the density contour expands dramatically. Poulenc sits near Classical-era territory, whereas Schulhoff and Bart\'ok occupy positions of lower focus and high connection, as they start to move away from tonal constructs. This progressive fanning-out aligns with standard music-historical narratives of increasing chromaticism and stylistic individuation \cite{schoenberg1975}.\footnote{Detailed composer-level distributions appear in supplementary material, Appendix~A.}

\section{Discussion}

\subsection{Coherence as a Multi-Dimensional Achievement}

Our findings demonstrate that tonal coherence is not a unidimensional phenomenon but arises through multiple organizational strategies that different traditions weight differently. The two dimensions we identify---tonal focus and tonal connection---resonate with the longstanding theoretical distinction between functional and transformational approaches to tonality.

The properties of classical music in these dimensions (high tonal connection with hierarchically differentiated interval weights) aligns with Neo-Riemannian theory. Coherence is achieved through systematic transformations that maintain intelligibility even as functional centers become obscured \cite{cohn2012audacious}. The weight analysis (Section~\ref{sec:dimensions}) provides supporting evidence; the dominance of perfect fifths combined with graded subordinate intervals is a necessary condition for voice-leading-based organization, though our static analysis cannot directly confirm temporal practices.

In contrast, the properties of popular music (high tonal focus with flatter intervallic organization) align with what Tagg \cite{tagg2014everyday} calls ``assertional'' tonality. Coherence is achieved through metrical emphasis, repetition, and gravitational anchoring, rather than transformational syntax. The platykurtic weight distributions suggest coherence emerges through mechanisms not predicated on systematic intervallic relationships. This aligns with what Everett \cite{everett2004rock} describes as ``metric assertion'' and ``melodic goal-directedness.''

The fact that these dimensions are partially independent ($r^2 < 0.04$) is in itself an interesting finding. Tonal focus and connection represent genuinely distinct compositional choices that interact but do not determine each other. The existence of boundary-crossing examples (Bach suites with maximal focus, pop songs with high tonal connection) demonstrates that genre-typical strategies are conventions, not constraints on what is musically possible.

\subsection{Implications for Computational Models}

Current music generation systems typically treat genre as a categorical variable. Our two-axis framework suggests a more nuanced approach. Rather than categorical templates, systems could navigate continuous trade-offs between tonal focus and connection. A ``focus'' parameter adjusting concentration near the tonic and a ``connection'' parameter adjusting the extent of tonal exploration would provide interpretable, continuous controls grounded in music-theoretic concepts.

For analytical applications, our decomposition reveals that single measures of ``tonalness'' or ``complexity'' conflate distinct dimensions. Pitch restriction indices rank popular music as ``simpler,'' while entropy-based measures rank classical as ``more complex.'' Both are accurate but incomplete---they capture only one axis of a two-dimensional space.

More broadly, our findings support the view that music-theoretic distinctions between functional and transformational approaches may reflect genuine cognitive or compositional realities rather than merely alternative analytical vocabularies.

\subsection{Limitations and Future Work}
\label{sec:robustness}

Our results are robust to the main methodological choices: the tonal focus difference persists across all threshold values from $k=2$ to $k=7$, the partial independence between dimensions holds at every threshold, key estimation produces realistic distributions, tonal connection estimates remain stable under key perturbation, weight structure metrics are not confounded by modulation rates, and a 12-dimensional reanalysis confirms that enharmonic spelling noise attenuates rather than inflates the observed differences.

Nevertheless, several limitations remain. Our analysis operates on static pitch-class distributions and cannot directly assess temporal practices such as voice-leading or harmonic progressions. Classical music's balanced weight profiles are consistent with voice-leading-based organization but could arise through other means. The binary classical/popular distinction obscures within-genre diversity, MIDI files lack ground-truth pitch spellings, and percussion tracks introduce noise, though filtering and normalization procedures mitigate these concerns.

These limitations suggest natural extensions. Sequential models, voice-leading efficiency metrics \cite{tymoczko2011} or path inference over note sequences, could test whether aggregate weight signatures correspond to actual voice-leading practice. Furthermore, listener studies connecting our measures to perceived coherence ratings would validate the framework perceptually and guard against normative interpretation of the organizational differences we identify.

\section{Conclusion}

This paper introduced a two-dimensional framework for analyzing tonal coherence, distinguishing \emph{tonal focus}---concentration near the tonic---from \emph{tonal connection}---structured intervallic relationships, which build on the Tonal Diffusion Model \cite{lieck2020tonal}. Applying this framework to 1,280 classical and 1,569 popular pieces, our findings indicate that these two musical traditions achieve coherence through different weightings of the same two dimensions rather than through fundamentally different mechanisms, with the partial independence between dimensions confirming that they represent genuinely distinct organizational choices.

When comparing Western classical music (1680--1920) with popular music (1950--2020), we find that they occupy overlapping yet distinguishable regions of the two-dimensional space. The clearest differentiation lies in how each tradition organizes pitch; classical music relies more heavily on perfect fifth relationships with hierarchically differentiated interval weights, popular music shows greater variance across pieces and genres. 

Our framework offers practical tools for computational musicology, \textit{i.e.}, continuous, interpretable parameters that replace categorical genre labels in both analysis and generation. 
%

\section{Acknowledgements}

This research has received partial funding by the Swiss National Science Foundation through the grants ``Distant listening: Transitions of Tonality'' (grant no. 215701) and ``Towards a Unified Model of Musical Form: Bridging Music Theory, Digital Corpus Research, and Computation'' (grant no. 10000183). 
We thank Mr. Claude Latour for generously supporting this research through the Latour chair
in digital musicology.
The authors would like to thank Uri Rom for his feedback on this paper.

\bibliography{smc2026bib}

\end{document}